\documentclass[english,11pt]{article}

\usepackage{amsfonts,amssymb,dsfont}
\usepackage{type1cm,graphicx}
\usepackage[usenames]{color}
\usepackage{bbm,mathtools}
\usepackage[left=2.55cm,top=2.55cm,right=2.55cm,bottom=2.55cm]{geometry}
\usepackage{amsthm}
\usepackage{multirow}
\usepackage[table,xcdraw]{xcolor}

\usepackage{graphicx}
\usepackage{algorithm}
\usepackage{algpseudocode}
\usepackage{dcolumn}
\usepackage{bm}
\usepackage{epstopdf}

\usepackage{hyperref}

\usepackage{extarrows}
\usepackage{mathtools}

\usepackage{xcolor} 
\usepackage{enumitem} 

\newcommand{\bk}[1]{\left |#1 \right \rangle}

\newcommand{\ketbra}[2]{|#1\rangle\! \langle #2|}

\newcommand{\nrm}[1]{\| #1 \|_{2}}
\newcommand{\bigO}[1]{\mathcal{O}\left( #1 \right)}

\newcommand{\tn}[1]{\textnormal{#1}}
\newcommand{\e}{\vec{e}}
\newcommand{\s}{\vec{s}}
\newcommand{\x}{\vec{x}}

\renewcommand\vec{\mathbf}

\newcommand{\flr}[1]{\lfloor #1 \rfloor}
\newcommand{\hf}{\stackrel{\$}{\longleftarrow}}

\newcommand{\N}{\mathbb{N}}
\newcommand{\R}{\mathbb{R}}
\newcommand{\Z}{\mathbb{Z}}
\newcommand{\I}{\mathbb{I}}

\newcommand{\mi}{\mathrm{i}}

\newcommand{\va}{\vec{a}}
\newcommand{\vb}{\vec{b}}
\newcommand{\vs}{\vec{s}}
\newcommand{\rd}[1]{\lfloor #1 \rceil}
\newcommand{\epi}[1]{e^{2\pi I  #1 }}
\newcommand{\tr}{\tilde{r}}

\newcommand{\tf}{\tilde{f}}

\newcommand{\tm}{\tilde{m}}
\newcommand{\tL}{\tilde{L}}

\newcommand{\vx}{\vec{x}}
\newcommand{\tb}{\textbf}
\newcommand{\tl}{\tilde{l}}
\newcommand{\tbl}[1]{\hspace{-18pt}\textbf{#1}}
\newcommand{\pol}[1]{\textnormal{polylog}{(#1) }}

\newcommand{\beq}{\begin{equation}}
\newcommand{\eeq}{\end{equation}}

 \newtheorem{thm}{Theorem}
  \newtheorem{prop}[thm]{Proposition}
 
 \newtheorem{lem}[thm]{Lemma}
 \newtheorem{defn}[thm]{Definition}

\numberwithin{thm}{section}

\numberwithin{equation}{section}

\makeatletter
\newenvironment{breakablealgorithm}
  {
   \begin{center}
     \refstepcounter{algorithm}
     \hrule height.8pt depth0pt \kern2pt
     \renewcommand{\caption}[2][\relax]{
       {\raggedright\textbf{\ALG@name~\thealgorithm} ##2\par}%
       \ifx\relax##1\relax 
         \addcontentsline{loa}{algorithm}{\protect\numberline{\thealgorithm}##2}%
       \else 
         \addcontentsline{loa}{algorithm}{\protect\numberline{\thealgorithm}##1}%
       \fi
       \kern2pt\hrule\kern2pt
     }
  }{
     \kern2pt\hrule\relax
   \end{center}
  }



\usepackage{enumitem}
\usepackage{mdframed}
\usepackage{xcolor}

\newmdenv[
  backgroundcolor=gray!20,
  topline=false,
  bottomline=false,
  rightline=false,
  leftline=false,
  skipabove=\topsep,
  skipbelow=\topsep
]{grayenumerate}

\begin{document}

\date{}

\author{Guangsheng Ma\footnote{Supported by China National Key Research and Development Projects 2025YFA1017200, the Fundamental Research Funds for the Central Universities (2025MS076, 2026SL010), BJNSF(no. 1242013); School of Mathematics and Physics, North China Electric Power University, Beijing, China. Email: 50902708@ncepu.edu.cn.}
\quad \quad \quad Hongbo Li\footnote{Corresponding author. Academy of Mathematics and Systems Science, Chinese Academy of Sciences; University of Chinese Academy of Sciences, Beijing, China. Email: hli@mmrc.iss.ac.cn.}}

\title{Quantum Blind Rotation for Fast Functional Bootstrapping}

\maketitle


\begin{abstract}
Fully homomorphic encryption (FHE) enables privacy-preserving cloud computation, but its efficiency is often limited by the cost of bootstrapping. In particular, existing functional bootstrapping techniques have complexity exponential in the plaintext size. In this work, we show that employing a single quantum server can reduce this dependence.

We propose a quantum functional bootstrapping algorithm that allows to evaluate any efficiently computable function in time polynomial in the plaintext size. For general functional bootstrapping over $l$-bit plaintexts, we obtain a time--space tradeoff: poly($l$)-time evaluation can be achieved with O$(2^l)$ qubits, while reducing the space complexity increases the time complexity.

Technically, we extend a key classical cryptographic operation, known as \emph{blind rotation}, to the quantum setting by replacing polynomial-exponent encoding with quantum phase encoding. Underlying our extension are insights for the quantum extension of polynomial-based cryptographic tools that may gain dramatic speedups.

\end{abstract}

\section{Introduction}

Fully Homomorphic Encryption (FHE) is an encryption scheme that supports direct computation on encrypted data. In cloud computing environments, the server performs homomorphic operations on FHE-encrypted data, allowing the outsourcing of computations without revealing the data. A central difficulty in FHE is that the noise contained in a ciphertext grows during homomorphic computations and may eventually exceed the decryption margin. To support unlimited level of homomorphic computations, Gentry proposed the revolutionary idea of \emph{bootstrapping} \cite{gentry2009fully1}, which allows control of error growth. Over the past decade, a large number of literature is devoted to explore and develop bootstrapping methods \cite{gentry2011implementing,biasse2015fhew,hiromasa2016packing,chillotti2017faster,halevi2021bootstrapping,micciancio2021bootstrapping}. Despite significant progress, bootstrapping remains one of the most expensive operations in FHE crypto-systems, and its optimization continues to receive considerable attention \cite{liu2023amortized,de2024faster,ma2024fast,bae2024plaintext}.

The main idea of bootstrapping is to homomorphically evaluate the decryption circuit. For the phase $qI+m+e$ in encrypted data, where the plaintext $m$ is sandwiched by the head error $qI$ and tail error $e$, the decryption circuit recovers plaintext $m$ via modulo-$q$ to remove the head error, followed by rounding operation to remove the tail error $e$. By decrypting within an encrypted environment, bootstrapping generates a new ciphertext encrypting the same plaintext but with much smaller tail error.

\emph{Functional bootstrapping}, first introduced in third-generation FHE schemes (FHEW/TFHE \cite{ducas2015fhew,chillotti2017faster}), extends this procedure by simultaneously refreshing a ciphertext and evaluating a prescribed function on its plaintext. It plays a crucial role in privacy-preserving machine-learning \cite{chillotti2021programmable}. The core technique behind functional bootstrapping is known as \emph{blind rotation} \cite{chillotti2017faster}, which evaluates the decryption circuit over the polynomial ring $\Z[x]/(x^N+1)$. Blind rotation first changes the modulo $q$ to $2N$ and lifts the phase $qI+m+e$ in encrypted data to exponent $x^{m+e}$, where the head error disappeared by polynomial modular reduction, and the tail error can be easily separated in the exponent by $x^{m+e}=x^{m}x^{e}$. Then, to bring down plaintext $m$ from the exponent to the coefficient, a test polynomial is used, which is essentially the look-up table of a prescribed function $f$, so that when $m$ is brought down to the coefficient, the error factor can be removed, and the encrypted plaintext becomes $f(m)$.

This construction creates a fundamental representation cost in functional bootstrapping. If $m$ is an $l$-bit plaintext, then its domain contains $L=2^l$ possible values. A generic lookup-table representation must distinguish all $L$ possible values, so that the degree of the test polynomial grows exponentially with $l$. Consequently, a direct implementation of functional bootstrapping has a complexity exponential in the plaintext-size.

Improving the efficiency of functional bootstrapping has became an active area of research \cite{chillotti2017improving,carpov2019new,kluczniak2021fdfb,micciancio2018ring,micciancio2021bootstrapping,yang2021tota,liu2023batch2}. Several works exploit parallelism or batching methods \cite{liu2023batch,liu2023batch2,micciancio2021bootstrapping,de2024faster}. The underlying observation is that to ensure the security, the parameters that even encrypting a single-bit are often large enough to support the simultaneous processing of several low-precision plaintexts. This idea can be used to reduce the amortized cost of functional bootstrapping over multiple single-bit plaintexts. However, it does not address the cost of bootstrapping a single large-precision plaintext.

To handle large-precision plaintexts, segmented bootstrapping strategy \cite{liu2022large,yang2021tota,ma2024fast} split long plaintexts into shorter blocks, bootstrap each block sequentially, and finally recombine all the blocks. This strategy achieves a complexity polynomial in plaintext size for evaluating simple functions (e.g., linear functions and most-significant-bit extraction). For a general nonlinear function $f$, however, recombining ciphertext blocks Enc$(f(m_i))$ into a single ciphertext Enc$(f(\sum_i m_i))$ is generally difficult. To the best of our knowledge, for general-purpose functional bootstrapping, the complexity remains exponential in plaintext-size.

\tbl{Motivation.} The above cost mainly arises form that classical functional bootstrapping relies heavily on polynomial representations. In contrast, quantum computation provides alternative ways of representing and processing information through qubit amplitudes and computational-basis states. Such quantum representations and primitives have enabled computational speedups in various problems, including lattice-based cryptanalysis \cite{cramer2016recovering,cramer2017short}, machine learning \cite{biamonte2017quantum}, and linear-system solving \cite{harrow2009quantum}. This motivates us to ask:

\vspace{0.3cm}
\emph{Can a single quantum server be used to improve the efficiency of functional bootstrapping? }
\vspace{0.3cm}

We study this question in a hybrid quantum-classical cloud-computing model, in which the client remains entirely classical and communicates with a single quantum server using only classical channels. This scenario aligns with the current work mode of quantum cloud platforms (e.g., Google Quantum AI), avoiding the challenges of quantum communication (i.e., long-distance qubit transmission) and the risks of collusion in a multi-server setting.
At the same time, achieving a quantum advantage under such restrictions is particularly challenging: the client has no quantum capability, and neither quantum communication nor pre-shared entanglement is available.

\begin{figure*}
  \includegraphics[width=1\textwidth]{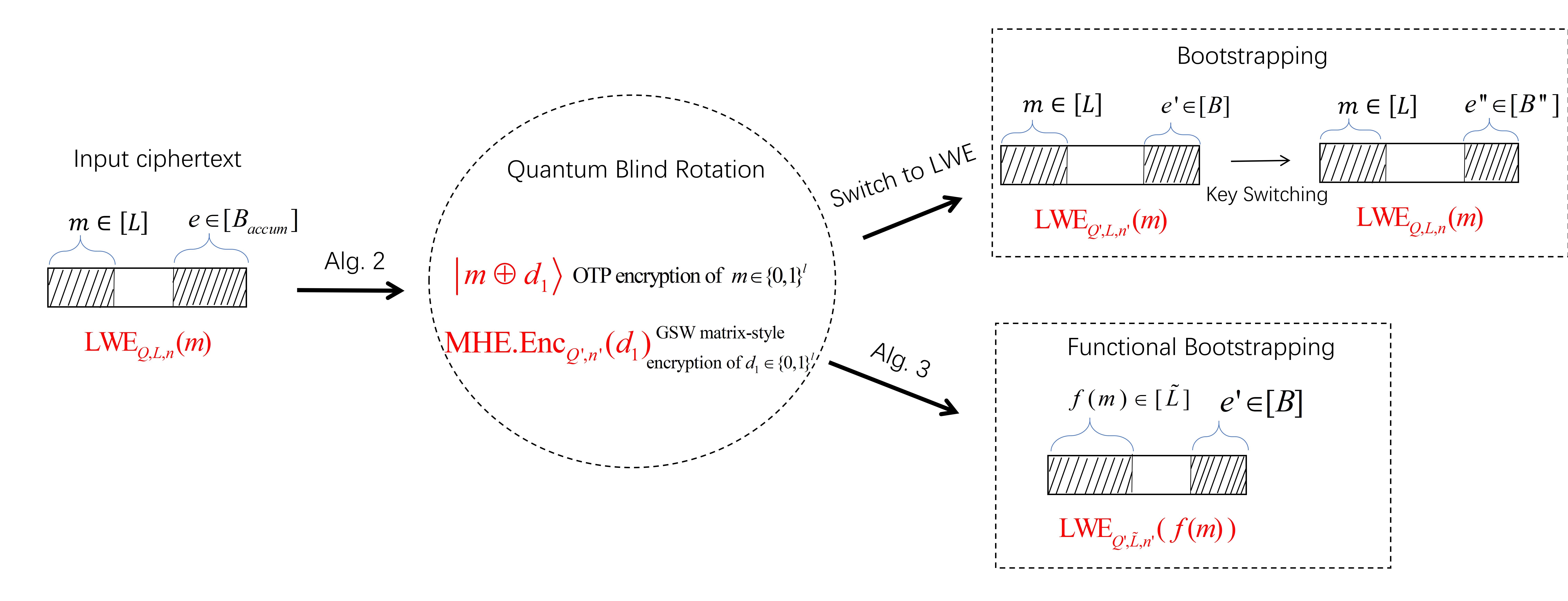}
  \caption{Quantum fast bootstrapping and functional bootstrapping. An input LWE encryption, after quantum blind rotation (Algorithm \ref{123}), can be converted into a combination of quantum and classical ciphertexts. These ciphertexts can then be merged into an encryption of either the original input message (Subsection \ref{101}) or its function value (Algorithm \ref{909}). When setting $L'=L$ and $m'=m$, the output LWE ciphertext has parameters
  consistent with those of the input.}\label{6}
\end{figure*}

\tbl{Our contribution.} With quantum resources limited to only a quantum server, we construct a quantum functional-bootstrapping algorithm whose time dependence on the plaintext size is polynomial for efficiently computable functions (cf. Theorem \ref{4.4}). For arbitrary functions, polynomial-time evaluation is possible under a QRAM assumption. Our construction has three main components: a quantum analogue of blind rotation (Algorithm \ref{123}), a hybrid quantum--classical procedure for function evaluation (Algorithm \ref{909}), and a ciphertext-format switching procedure (Subsection \ref{101}). Figure~\ref{6} summarizes the overall quantum functional-bootstrapping procedure.\\

\tbl{Quantum blind rotation.} The main technical contribution is called quantum blind rotation (Algorithm \ref{123}), which reduces the computational complexity by utilizing qubits's amplitudes to replace polynomial's exponents in classical blind rotation.

At a high level, quantum blind rotation homomorphically computes the LWE-phase $qI+m+e$ in the quantum phase. The head error $qI$ is removed by the natural 2$\pi$-periodicity of phase factor, and the tail error $e$ can be removed by quantum measurements at a later stage. The first challenge arises in extracting the large-precision plaintext from quantum phase into computational-basis states via quantum phase estimation.

Quantum phase estimation requires repeated oracle queries to implement phase rotations with angles of increasing precision. Its (query) complexity scales exponentially with the desired number of bit of precision, i.e., the number of qubits in the resulting computational-basis states. Fortunately, by encryption conditional rotation technique, we can generates states with different phases angles using the same encryption, enabling efficient phase estimation with complexity that scales only polynomially with the plaintext-size.

After blind rotation, the resulting state is protected by a quantum Pauli one-time pad, while the corresponding Pauli keys remain encrypted under the classical FHE scheme. For ordinary bootstrapping, these quantum and classical ciphertext components can be combined to obtain a ciphertext in the same format as the input.

For functional bootstrapping, the remaining task is to evaluate the desired function on the encrypted data. Specifically, for a function $f:\Z_{m}\rightarrow\Z_{\tm}$, the task is to homomorphically evaluate the unitary operator $U: \bk{m}\bk{0}\rightarrow \bk{m}\bk{f(m)}$ over quantum encryption of $\bk{m}$. This task is reminiscent of private information retrieval (PIR), a multi-party secure computation protocol that allows clients to retrieve specific data DB$(i)$ from a database without revealing the query $i$.

Existing PIR and quantum PIR (QPIR) protocols exhibit different tradeoffs among communication, preprocessing, and security. Information-theoretically secure classical PIR requires at least linear communication complexity \cite{chor1998private}, and this lower bound also hold in the quantum setting \cite{baumeler2015quantum}. Computationally secure PIRs allows lower communication and online complexity. A notable example in classical PIR is the DEPIR protocol proposed by Lin et al. \cite{lin2023doubly}, which achieves logarithmic online time and communication complexity, significantly improving upon earlier sub-linear complexity \cite{corrigan2020private,corrigan2022single}. However, its practical performance can be less favorable than that of asymptotically slower schemes \cite{zhou2024piano}, and its security relies on Ring-LWE with a super-polynomial noise (modulus-to-noise) ratio. This is weaker than the gold standard LWE security with a polynomial noise ratio. Designing PIR protocols that improves efficiency and security remains an active area of research.

Quantum cloud computing also offer a way to reduce PIR communication. Representative works include Le Gall's QPIR protocol \cite{le2012quantum} with a communication complexity of O$(\sqrt{N})$, and Kerenidis et al.'s protocol \cite{kerenidis2016information} with polylog$(N)$ complexity using pre-shared entanglement. However, most existing QPIR protocols \cite{song2020capacity,allaix2021high,song2021quantum} rely on client-side quantum computing, quantum communication, or multiple non-colluding servers. These requirements are incompatible with the cloud-computing model considered here.

\tbl{Hybrid quantum-classical PIR (QCPIR).} As a building block, we introduce a QPIR framework that integrates the quantum fully homomorphic encryption (QFHE) with quantum random access memory (QRAM) circuits. While constructing PIR from (quantum) FHE is not new in literature, this straightforward approach leads to a QPIR protocol that remains of independent interest, when compared to advanced DEPIR and previous QPIRs. Table~\ref{61} compares its resource requirements with existing classical and quantum PIR protocols.\\

\begin{table*}[t]
\resizebox{\textwidth}{!}{
\begin{tabular}{llccccclclcll}
                                  &           & \multicolumn{2}{c}{\textbf{Time}}                       &                      & \multicolumn{2}{c}{\textbf{Communication}}                    &  &                           &  &                         &  & \multicolumn{1}{c}{}                                                                                                                                                                                                \\ \cline{3-4} \cline{6-7}
                                  &           & Online                             & Offline              &                      & Complexity              & Round                &  & \textbf{Space Complexity}         &  & \textbf{ Client-Side Computation} &  & \multicolumn{1}{c}{\textbf{Security Assumption}}                                                                                                                                                                                        \\ \hline
\multicolumn{2}{l}{\textbf{Classical Single-Server}} &                                    &                      &                      &                               &                      &  & classical bits            &  &                         &  &                                                                                                                                                                                                                     \\
                                  &           & \multicolumn{1}{l}{}               & \multicolumn{1}{l}{} & \multicolumn{1}{l}{} & \multicolumn{1}{l}{}          & \multicolumn{1}{l}{} &  & \multicolumn{1}{l}{}      &  & \multicolumn{1}{l}{}    &  &                                                                                                                                                                                                                     \\
{\cite{boneh2017constraining}}                       &           & $N$                                & $N$                  &                      & polylog $N$                   & 1                    &  & $N$                       &  & log $N$                 &  & \multicolumn{1}{c}{PRFs}                                                                                                                                                                                            \\
                                  &           & \multicolumn{1}{l}{}               & \multicolumn{1}{l}{} & \multicolumn{1}{l}{} & \multicolumn{1}{l}{}          & \multicolumn{1}{l}{} &  & \multicolumn{1}{l}{}      &  & \multicolumn{1}{l}{}    &  &                                                                                                                                                                                                                     \\
{\cite{corrigan2022single}}                        &           & $N^{1/2}$                              &   $N$         &                      & $N^{1/2}$                     & 1                    &  & $N$                       &  & $0$                      &  & \multicolumn{1}{c}{OWF}                                                                                                                                                                                             \\
                                  &           & \multicolumn{1}{l}{}               & \multicolumn{1}{l}{} & \multicolumn{1}{l}{} & \multicolumn{1}{l}{}          & \multicolumn{1}{l}{} &  & \multicolumn{1}{l}{}      &  & \multicolumn{1}{l}{}    &  &                                                                                                                                                                                                                     \\
\rowcolor[HTML]{FFFE65}
{\cite{lin2023doubly}}                       &           & {\color[HTML]{FE0000} polylog $N$}   & $N^{1+\epsilon}$     &                      & polylog $N$                   & 1                    &  & $N^{1+\epsilon}$          &  & 0                       &  & \multicolumn{1}{c}{\cellcolor[HTML]{FFFE65}{\color[HTML]{FE0000} }}                                                                                                                                                 \\
\rowcolor[HTML]{FFFE65}
                                  &           & $N^{o(1)}$                         & $N^{1+o(1)}$         &                      & $N^{o(1)}$                    & 1                    &  & $N^{1+o(1)}$              &  & 0                       &  & \multicolumn{1}{c}{\multirow{-2}{*}{\cellcolor[HTML]{FFFE65}{\color[HTML]{FE0000} \begin{tabular}[c]{@{}c@{}} RingLWE with Quasi-polynomial Noise Ratio\\ RingLWE with Sub-sub-exponential Noise Ratio \end{tabular}}}} \\
                                  &           & \multicolumn{1}{l}{}               & \multicolumn{1}{l}{} & \multicolumn{1}{l}{} & \multicolumn{1}{l}{}          & \multicolumn{1}{l}{} &  & \multicolumn{1}{l}{}      &  & \multicolumn{1}{l}{}    &  &                                                                                                                                                                                                                     \\ \hline
\textbf{Quantum Single-Server}           &           &                                    &                      &                      &                               &                      &  &                           &  &                         &  &                                                                                                                                                                                                                     \\
                                  &           & \multicolumn{1}{l}{}               & \multicolumn{1}{l}{} & \multicolumn{1}{l}{} & \multicolumn{1}{l}{}          & \multicolumn{1}{l}{} &  & \multicolumn{1}{l}{}      &  & \multicolumn{1}{l}{}    &  &                                                                                                                                                                                                                     \\
{\cite{le2012quantum}}                      &           & $\sqrt{N}$ & 0                    &                      & $\sqrt{N}$ (Quantum Comm.)       & 2              &  & $\sqrt{N}$ shared entanglement &  & $\sqrt{N}$ quantum gates       &  & \multicolumn{1}{c}{Anchored Privacy (Weaker than ITS)}                                                                                                                                                              \\
                                  &           & \multicolumn{1}{l}{}               & \multicolumn{1}{l}{} & \multicolumn{1}{l}{} & \multicolumn{1}{l}{}          & \multicolumn{1}{l}{} &  & \multicolumn{1}{l}{}      &  & \multicolumn{1}{l}{}    &  &                                                                                                                                                                                                                     \\
{\cite{kerenidis2016information}}                      &           & {\color[HTML]{FE0000} polylog $N$} & 0                    &                      & log $N$ (Quantum Comm.)       & log $N$              &  & $N $ shared entanglement &  & $N$ quantum gates       &  & \multicolumn{1}{c}{Anchored Privacy}                                                                                                                                                              \\
                                  &           & \multicolumn{1}{l}{}               & \multicolumn{1}{l}{} & \multicolumn{1}{l}{} & \multicolumn{1}{l}{}          & \multicolumn{1}{l}{} &  & \multicolumn{1}{l}{}      &  & \multicolumn{1}{l}{}    &  &                                                                                                                                                                                                                     \\
{\cite{song2021quantum}}                      &           & {\color[HTML]{FE0000} polylog $N$} & 0                    &                      & log $N$ (Quantum Comm.)       & log $N$              &  & $N $ shared entanglement &  & $N$ quantum gates       &  & \multicolumn{1}{c}{Honest Server Model}                                                                                                                                                              \\
                                  &           & \multicolumn{1}{l}{}               & \multicolumn{1}{l}{} & \multicolumn{1}{l}{} & \multicolumn{1}{l}{}          & \multicolumn{1}{l}{} &  & \multicolumn{1}{l}{}      &  & \multicolumn{1}{l}{}    &  &                                                                                                                                                                                                                     \\
\rowcolor[HTML]{FFFC9E}
This work                         &           & {\color[HTML]{FE0000} polylog $N$} & 0                    &                      & polylog $N$ (Classical Comm.) & 1                    &  & $N$ qubits                &  & {\color[HTML]{FE0000} 0 quantum gate}                       &  & \multicolumn{1}{c}{\cellcolor[HTML]{FFFC9E}{\color[HTML]{FE0000} LWE with Noise Ratio poly$(\lambda)$}}                                                                                                      \\
\cellcolor[HTML]{FFFFFF}          &           & \multicolumn{1}{l}{}               & \multicolumn{1}{l}{} & \multicolumn{1}{l}{} & \multicolumn{1}{l}{}          & \multicolumn{1}{l}{} &  & \multicolumn{1}{l}{}      &  & \multicolumn{1}{l}{}    &  &                                                                                                                                                                                                                     \\ \hline\textbf{Lower Bound for QPIR with Single-Server}           &           &                                    &                      &                      &                               &                      &  &                           &  &                         &  &                                                                                                                                                                                                                     \\
                                  &           & \multicolumn{1}{l}{}               & \multicolumn{1}{l}{} & \multicolumn{1}{l}{} & \multicolumn{1}{l}{}          & \multicolumn{1}{l}{} &  & \multicolumn{1}{l}{}      &  & \multicolumn{1}{l}{}    &  &                                                                                                                                                                                                                     \\
{\cite{baumeler2015quantum,aharonov2019quantum}}                      &           & Unknown & Unknown                &                      & $N$       & Unknown             &  & Unknown &  &Unknown      &  & \multicolumn{1}{c}{Information-Theoretic Privacy (Specious Model)}                                                                                                                                                              \\
                                  &           & \multicolumn{1}{l}{}               & \multicolumn{1}{l}{} & \multicolumn{1}{l}{} & \multicolumn{1}{l}{}          & \multicolumn{1}{l}{} &  & \multicolumn{1}{l}{}      &  & \multicolumn{1}{l}{}    &  &                                                                                                                                                                                                                     \\

\end{tabular}}
\caption{Comparison of classical and quantum single-server PIR protocols. All columns (except the last ``Security Assumption'' one) omit poly($\lambda$) factors, for security parameter $\lambda$, constant factors, and  unimportant polylog($N$) factors. Here, $\epsilon > 0$ is an arbitrarily small constant. \label{61}}
\end{table*}

\emph{Compared to classical PIR.} Compared with the DEPIR scheme of Lin et al. \cite{lin2023doubly}, which also achieves polylogarithmic online time complexity, the QCPIR has the following advantages:
\begin{itemize}
\item[] \emph{Stronger Security.} When instantiated with Brakerski's QFHE scheme \cite{brakerski2018quantum}, its security is based on standard LWE with polynomial noise ratio, which is stronger than the super-polynomial RLWE security of classical DEPIR schemes \footnote{ The LWE problem with polynomial noise ratio is widely regarded as the gold standard for LWE-based security. First, as the noise ratio decreases, the LWE problem becomes harder \cite{regev2009lattices}. Second, Ring-LWE, tied to lattices of special structures, is assumed to be less secure than the general lattice-based LWE at comparable (up to polynomial) noise ratios. Third, polynomial-time quantum (and classical) algorithms for ideal lattice problems with sub-exponential factors are (reportedly) known \cite{cramer2016recovering,cramer2017short,biasse2022mildly}, leading to attacks on sub-exponential RLWE over certain rings, such as cyclotomic polynomial ring.}.

\item[] \emph{No Preprocessing.} The QCPIR protocol requires no precomputation, while DEPIR relies on precomputing and storing a table of size O$(N^{1+\epsilon})$ to support efficient online-search of encrypted indices \cite{lin2023doubly}.

\item[] \emph{Lower Space Complexity.} The QCPIR protocol uses O$(N)$ qubits in general case, and this space complexity can be reduced to polylog$(N)$ when the database is efficiently computable, or its quantum preparation unitary (\ref{60}) is efficiently implementable. By comparison, the classical DEPIR protocol requires O$(N^{1+\epsilon})$ bits of classical memory.
\end{itemize}

\emph{Compared to quantum PIR.} The QCPIR scheme supports an entirely classical client and requires only one round of classical communication with polylog($N$) communication complexity. In contrast, previous QPIR schemes typically require O(log$N$) rounds of quantum communication or O($\sqrt{N}$) communication complexity.\\

\tbl{Quantum functional bootstrapping.} Combining quantum blind rotation, QCPIR, and ciphertext format switching yields a quantum functional bootstrapping algorithm with several selective components and time-space tradeoffs. Given an encryption of an $l$-bit plaintext $m\in[2^l]$, the algorithm supports the following two settings:
\begin{itemize}
\setlength{\leftskip}{0pt}
\item[] \emph{Efficiently computable functions.} For a function $f(m):\Z_{2^l}\mapsto\Z_{2^{\tl}}$ that can be efficiently computed in classical time poly$(l,\tl)$, the algorithm outputs the encryption of $f(m)$ in comparable time poly$(l,\tl)$. This enables the fast evaluation of certain non-linear and non-trivial functions, such as $f(m)=e^m$ and $f(m)=\sqrt{m}$, that are not naturally handled by existing functional bootstrapping with segmented strategies.

\item[] \emph{General functions.} For a general function $f:\Z_{2^l}\mapsto\Z_{2^{\tl}}$, the algorithm can produce the encryption of $f(m)$ with either a runtime of poly$(l,\tl)$ and O$(2^l)$ ancilla qubits, or a runtime of poly$(2^l,\tl)$ and O$(1)$ ancilla qubits. Additional tradeoffs between time and space overheads are also available.
\end{itemize}

Intuitively, the improved complexity over classical polynomial-based functional bootstrapping is partly due to the ability to control precision while converting quantum phases into computational-basis states through phase estimation. Unlike classical polynomials, which enforce a rigid one-to-one correspondence between coefficients and exponents, quantum phase offer more flexible precision management.

\tbl{Future work.}
Our quantum algorithm for functional bootstrapping has polynomial time dependence on plaintext-size $l$, while its dependence on the ciphertext dimension $n$ remains linear. Recent advances in parallel bootstrapping \cite{liu2023batch2}, relying on RLWE and FFT techniques, have exponentially reduced the dependence on $n$, yielding an amortized complexity of $\tn{poly} (\log n, N) $, where $N>2^l$. It would be valuable to explore whether more algebraic approach they used, such as Chinese Remainder Theorem and parallel RLWE bootstrapping techniques, can be incorporated into the quantum framework. This integration may lead to improved amortized complexity, possibly $\tn{poly} (\log n, l)$.

We emphasize that our result does not establish a formal separation between quantum blind rotation and classical algorithms. It remains a question whether the quantum blind rotation can be efficiently simulated classically.

\section{Preliminary}
\textbf{Notation.} For any $q\in\N$, let $\Z_q$ be the ring of integers modulo $q$, and let $[q]$ be the subset of integer $\{0,1,...,n-1\}$. For any $a\in \R$, let $\lfloor a \rceil$ denote integer closet to $a$, with $\lfloor 1/2 \rceil=1$, and let $\lfloor a \rceil_{t}$ denote the number in $Z_t$ closet to $a$.

The $L^2$-norm of vector $\vec{a}=(a_j)$ is defined as $\nrm{\vec{a}}:=\sqrt{\sum_{j} |a_j|^2}$. We use $\I$ to denote the identity matrix whose size is obvious from the context.

The \emph{Gaussian density function} is defined as $\rho_s(\vx):=e^{-\pi(||\vx||_2/s)^2}$, where $\vx\in\R^{n}$. The \emph{discrete Gaussian density function} $D_{\mathbb{Z}^{n}_q,B}$ is supported over the set $\{\vx \in \Z_q^{n} : ||\vx||_2<B\}$ such that Pr$[D_{\mathbb{Z}^{n}_q,B}=\vx]\propto\rho_s(\vx)$. We denote sampling from $D_{\mathbb{Z}^{n}_q,B}$ uniformly at random by $\vx \hf D_{\mathbb{Z}^{n}_q,B}$.

A function $f=f(\lambda)$ is \emph{negligible} in $\lambda$, if for any polynomial function $P(\lambda)$, it holds that $\displaystyle\lim_{\lambda\rightarrow\infty} f(\lambda)P(\lambda)=0$. A probability $p(\lambda)$ is \emph{overwhelming} if $1-p=\tn{negl}(\lambda)$

For a qubit system that has probability $p_i$ to be in state $\bk{\psi_i}$ for every $i$ in some index set, the \emph{density matrix} is defined by $\rho=\sum_{i}p_i\ketbra{\psi_i}{\psi_i}$.

\subsection{Classical Fully Homomorphic Encryption}
We base our definitions on the works \cite{brakerski2018quantum} and \cite{mahadev2018classical}.

\begin{defn}[Classical homomorphic encryption scheme] A homomorphic (public-key) encryption scheme HE = (HE.Keygen, HE.Enc, HE.Dec, HE.Eval) for single-bit plaintexts is a quadruple of PPT algorithms as following:
\begin{itemize}
\setlength{\leftskip}{0pt}
\item \tn{\tb{HE.KeyGen}}: The algorithm $(pk,evk,sk)\leftarrow$ \tn{HE.Keygen}$(1^{\lambda})$ on input the security parameter $\lambda$ outputs a public encryption key $pk$, a public evaluation key $evk$ and a secret decryption key $sk$.
\item \tn{\tb{HE.Enc}}: The algorithm $c\leftarrow$ \tn{HE.Enc}$_{pk}(\mu)$ takes as input the public key $pk$ and a single bit message $\mu\in\{0,1\}$, and outputs a ciphertext $c$. The notation HE.Enc$_{pk}(\mu;r)$ will be used to represent the encryption of message $\mu$ using random vector $r$.
\item \tn{\tb{HE.Dec}}: The algorithm $\mu^*\leftarrow$ \tn{HE.Dec}$_{sk}(c)$ takes as input the secret key $sk$ and a ciphertext $c$, and outputs a message $\mu^*\in\{0,1\}$.
\item \tn{\tb{HE.Eval}}: The algorithm $c_{f}\leftarrow$ \tn{HE.Eval}$_{evk}(f,c_1,\ldots,c_l)$ on input the evaluation key $evk$, a function $f:\{0,1\}^l\rightarrow\{0,1\}$ and $l$ ciphertexts $c_1,\ldots,c_l$, outputs a ciphertext $c_f$ satisfying:
\begin{equation*}
\mathrm{HE.Dec}_{sk}(c_f) = f(\mathrm{HE.Dec}_{sk}(c_1),\ldots,\mathrm{HE.Dec}_{sk}(c_l))
\end{equation*}
with overwhelming probability.
\end{itemize}
\end{defn}

\begin{defn}[Classical pure FHE and leveled FHE]
A homomorphic encryption scheme is compact if its decryption circuit is independent of the evaluated function. A compact scheme is (pure) fully homomorphic if it can evaluate any efficiently computable boolean function. A compact scheme is leveled fully homomorphic if it takes $1^L$ as additional input in key generation, where parameter $L$ is polynomial in the security parameter $\lambda$, and can evaluate all Boolean circuits of depth $\leq L$.
\end{defn}

A leveled classical FHE can be converted to into a pure classical FHE based on the circular security assumption \cite{gentry2009fully1,gentry2009fully}.

\subsection{Instance of Fully Homomorphic Encryption Based on the Learning With Errors}
\begin{defn}[Learning with errors (LWE) problem \cite{regev2009lattices}] Let $m,n,q$ be integers, and let $\chi$ be a distribution on $\Z_{q}$. The search version of LWE problem is to find $\s\in\Z_q^n$ given some LWE samples ($\vec{a}$, $\vec{a} \cdot \s+e$ mod $q$), where $\vec{a} \hf \Z_q^{ n}$ is sampled uniformly at random, and $e \leftarrow \chi$.
\end{defn}

Until now, there is no polynomial time algorithm that solves LWE problem for certain parameter regimes. In particular, Regev showed that a quantum polynomial-time algorithm for LWE problem with a certain polynomial noise ratio implies a quantum algorithm for gap shortest vector problem (SVP) with an approximation factor of $\sqrt{n}$. The later is a hard lattice problem in NP $\cap$ coNP, and LWE problem is generally considered secure even against quantum adversaries.

Based on LWE problem, several FHE schemes have been developed, with the major difference lying in where they story the plaintext. Notably,

\begin{defn}[Standard LWE encryption scheme]
The standard LWE encryption scheme stores the plaintext message in the high bits, as following:
\begin{itemize}
    \setlength{\leftskip}{0pt}  
\item \tn{\tb{LWE.KeyGen}}: Generate a matrix $\mathbf{\tilde {A}} \in \mathbb{Z}_q^{t \times n}$ uniformly at random. Choose $\mathbf{s} \in \mathbb{Z}_2^n$ \footnote{For simplicity, we consider using a binary secret key instead of one from $\Z^{n}_{q}$. This is a common setting in TFHE bootstrapping studies \cite{chillotti2020tfhe}, and most techniques developed in this context, like blind rotation, can extend to the non-binary case.} uniformly at random. Create error $\e\in\Z^{t}_q$ with each entry sampled from Gaussian distribution $D_{\mathbb{Z}_q,B}$. Compute the vector $\mathbf{\tilde {b}} = \mathbf{\tilde {A}}\mathbf{s} + \mathbf{e} \in \mathbb{Z}_q^t$. Select $l\in \Z$ as the plaintext size, and let $L=2^l$ be the plaintext space.

    The public key is $(\mathbf{\tilde {A}}, \mathbf{\tilde {b}})$ and the secret key is $\mathbf{s}$. The evaluation key are LWE encryptions for $s_i$, $s_is_j$, and their multiples of powers of two.

\item \tn{\tb{LWE.Enc}}$_{(\mathbf{\tilde {A}}, \mathbf{\tilde {b}})}(m)$: To encrypt a message $m \in [L]$, choose a random vector $\mathbf{x} \in \{0, 1\}^t$. Output $(\va,b) = (\mathbf{x}^T \mathbf{\tilde {A}}, \mathbf{x}^T \mathbf{\tilde {b}} + \lfloor q/L \rfloor \cdot m) \in \mathbb{Z}_q^{n+1}.$
\item \tn{\tb{LWE.Dec}}$_{\s}(\va,b)$: Output $\lfloor\frac{L}{q}(b-\vec{a} \cdot \s)\rceil_{L} \in \Z_L.$

\item  \tn{\tb{LWE.Add}}$((\va_1,b_1),(\va_2,b_2))$: Output $(\va_1+\va_2,b_1+b_2) \in \Z_q$

\item \tn{\tb{LWE.Mul}}$((\va_1,b_1),(\va_2,b_2))$: Compute $(\va_1\bigotimes \va_2, b_2\cdot\va_1+b_1\cdot\va_2, b_1b_2)\in Z^{n^2+n+1}_q$,\footnote{Given inputs \tn{LWE.Enc}$(m_1)=(\va_1,b_1)$ and \tn{LWE.Enc}$(m_2)=(\va_1,b_1)$, the result is an encryption of $m_1m_2$ under the secret key $(1,\s, \s\bigotimes \s)\in Z^{n^2+n+1}_2$.} and then utilize ``relinearization'' procedure\footnote{This paper does not relies on relinearization, so we omit the details. The basic idea of relinearization is similar to that of key switching, which is discussed in Section 5. We refer interested readers to \cite{brakerski2014efficient} for further information. } to reduce the ciphertext dimension to $n$.
\end{itemize}
\end{defn}

Homomorphic addition cause a linear increase in the ciphertext error, whereas homomorphic multiplication leads to exponential error growth. To ensure correct decryption, the accumulated noise must remain below the bound of $\flr{q/L}/2$. Bootstrapping that allows error resetting is therefore crucial. Detailed bootstrapping methods for LWE encryption can be found in \cite{brakerski2014efficient}. Functional bootstrapping and blind rotation technique are presented in \cite{chillotti2017faster}.

\subsection{Quantum Fully Homomorphic Encryption}
\begin{defn}[Quantum homomorphic encryption scheme]
A quantum homomorphic encryption (QHE) is a sequence of algorithms (QHE.Keygen, QHE.Enc,QHE.Dec, QHE.Eval). A hybrid framework of QHE with classical key generation is given below.
\begin{itemize}
   \setlength{\leftskip}{0pt}
\item \tn{\textbf{QHE.KeyGen}} The algorithm $(pk,evk,sk)\leftarrow$ \tn{HE.Keygen}$(1^{\lambda})$ takes as input the security parameter $\lambda$, and outputs a public encryption key $pk$, a public evaluation key $evk$, and a secret key $sk$.
\item \tn{\textbf{QHE.Enc}}  The algorithm $\bk{c}\leftarrow$ \tn{QHE.Enc}$_{pk}(\bk{m})$ takes the public key $pk$ and a single-qubit state $\bk{m}$, and outputs a quantum ciphertext $\bk{c}$.
\item \tn{\textbf{QHE.Dec}} The algorithm $\bk{m^*}\leftarrow$ \tn{QHE.Dec}$_{sk}(\bk{c})$ takes the secret key $sk$ and a quantum ciphertext $\bk{c}$, and outputs a single-qubit state $\bk{m^*}$ as the plaintext.
\item \tn{\textbf{QHE.Eval}}  The algorithm $\bk{c'_1},\ldots,\bk{c'_{l'}}\leftarrow$ \tn{QHE.Eval}$(evk,C,\bk{c_1},\ldots,\bk{c_l})$ takes the evaluation key $evk$, a classical description of a quantum circuit $C$ with $l$ input qubits and $l'$ output qubits, and a sequence of quantum ciphertexts $\bk{c_1},\ldots,\bk{c_{l}}$. Its output is a sequence of $l'$ quantum ciphertexts $\bk{c'_1},\ldots,\bk{c'_{l'}}$.
\end{itemize}
\end{defn}

Similar to the classical setting, the difference between leveled QFHE and pure QFHE is whether there is an a-priori bound on the depth of the evaluated circuit.

\subsection{Instance of Quantum Fully Homomorphic Encryption based on Pauli One-time Pad}
\begin{defn}[Pauli group and Clifford gates]
\item[]
The \textbf{Pauli group} on $n$-qubit system is $P_{n}=\{V_1 \otimes...\otimes V_n | V_j\in\{X,\ Z,\ Y,\ \I_2\}, 1\leq j\leq n \}$. The \emph{Clifford group} is the group of unitaries preserving the Pauli group:
$C_{n}=\{ V\in U_{2^{n}} |V P_n V{^\dagger}=P_n \}.$
A \textbf{Clifford gate} refers to any element in the Clifford group. A generating set of the Clifford group consists of the following gates:
 \begin{equation}\label{z4}
\begin{aligned}
X,& \quad Z, \quad
P = \left[ 
  \begin{array}{cc}
    1 & \\
     & \mi
  \end{array}
\right], \quad
H = \frac{1}{\sqrt{2}} \left[ 
  \begin{array}{cc}
    1 & 1 \\
    1 & -1
  \end{array}
\right],\
 \text{CNOT} = \left[ 
  \begin{array}{cccc}
    1 &  &  & \\
     & 1 &  & \\
     &  &  & 1 \\
     &  & 1 &
  \end{array}
\right].
\end{aligned}
\end{equation}
 \end{defn}
Adding any \emph{non-Clifford gate}, such as $T=\begin{bmatrix}1 & \\ & e^{\mi\frac{\pi}{4}}\end{bmatrix}$, to (\ref{z4}), leads to a universal set of quantum gates.

\begin{defn}[Pauli one-time pad encryption]
The Pauli one-time pad encryption, traced back to \cite{ambainis2000private}, encrypts a multi-qubit state qubitwise. The scheme for encrypting 1-qubit message $\bk{\psi}$ is as follows:
\begin{itemize}
   \setlength{\leftskip}{0pt}
\item[$\bullet$] \tn{\tb{POTP.KeyGen}}(). Sample two classical bits $a,b \leftarrow \{0,1\}$, and output $(a,b)$.
\item[$\bullet$] \tn{\tb{POTP.Enc}}$((a,b), \bk{\psi})$. Apply the Pauli operator $X^{a}Z^{b}$ to a 1-qubit state $\bk{\psi}$, and output the resulting state $\bk{\widetilde{\psi}}$.
\item[$\bullet$] \tn{\tb{POTP.Dec}}$((a,b),\bk{\widetilde{\psi}})$. Apply $X^{a}Z^{b}$ to $\bk{\widetilde{\psi}}$.
\end{itemize}
\end{defn}

\tbl{Information-theoretic security.} Note that, by $XZ=-ZX$, the decrypted ciphertext is the input plaintext up to a ignorable global phase factor $(-1)^{ab}$. In addition, the Pauli one-time pad encryption scheme guarantees the information-theoretic security, since for any 1-qubit state $\bk{\psi}$, it holds that
\begin{align}
\frac{1}{4}\sum_{a,b\in \{0,1\} }X^{a}Z^{b}|\psi\rangle\langle\psi|Z^{b}X^{a}=\frac{\I_2}{2}.
\end{align}

\tbl{Homomorphic Clifford gates.} In the QHE scheme based on Pauli one-time pad, e.g., \cite{broadbent2015quantum}, a 1-qubit state (plaintext) is encrypted in QOTP form $X^{a}Z^{b}\bk{\psi}$, and the Pauli keys $a,b\in\{0,1\}$ are also encrypted by using a classical FHE. The evaluation of any Clifford gate in this setting is easy:

\begin{itemize}
\setlength{\itemsep}{5pt}
 \setlength{\leftskip}{0pt}
\item[$\bullet$] \tb{POTP.EvalClifford}$(C, X^{a}Z^{b}\bk{\psi}, \tn{FHE.Enc}(a,b))$. To evaluate a Clifford gate $C$, directly apply $C$ on the $X^{a}Z^{b}\bk{\psi}$, and then homomorphically update $\tn{FHE.Enc}(a,b)$ according to the conjugate relation between gates $C$ and $XZ$ \footnote{For any Clifford gate $C$, we have $CX^{a}Z^{b} \bk{\psi}=X^{a'}Z^{b'}C\bk{\psi}$, where $a'$ and $b'$ can be represented as polynomials in $a,b$ of degree at most two. For example, the relation $HX=ZH$ leads to a key update of $a'=b$, $b'=a$ during the homomorphic evaluation of the Hadamard gate $H$.}.
\end{itemize}
After homomorphic evaluations, the outputs remain a Pauli-OTP encrypted state, together with the corresponding Pauli key in FHE-encrypted form.

\tbl{Homomorphic non-Clifford gates.} To evaluate of a non-Cliford gate, like T-gate, a dilemma arises: after applying the T-gate to an encrypted state, the relation $TX^{a}Z^{b}= P^{a}X^{a}Z^{b}T$ implies that the server needs to perform P$^a$ without knowing the value of $a$, but knowing its encryption. Mahadev \cite{mahadev2018classical} solved this dilemma by designing a matrix-style LWE-based FHE scheme, referred to as MHE, which is defined as follows:

\begin{itemize}
   \setlength{\leftskip}{0pt}
\setlength{\itemsep}{5pt}
\item[] \textbf{Quantum-capable MHE scheme (Scheme 5.2 in \cite{mahadev2018classical})}
\item[$\bullet$] \tb{MHE.KeyGen}: Choose $\e_{sk}\in\{0,1\}^{m}$ uniformly at random. Use \tn{GenTrap}$(1^n,1^m,q)$ in Theorem 5.1 of \cite{micciancio2012trapdoors} to generate a matrix $\*A$ $\in \mathbb{Z}^{m\times n}$, together with the trapdoor $t_{\*A}$. The secret key is $sk=(-\e_{sk},1) \in \mathbb{Z}^{m+1}_q$. The public key $\*A' \in \Z^{(m+1)\times n}_q$ is the matrix composed of $\*A$ (the first $m$ rows) and $\e^{T}_{sk}\*A \bmod q$ (the last row).

\item[$\bullet$] \tb{MHE.Enc}$_{pk}$($\mu$): To encrypt a bit $\mu\in\{0,1\}$, choose $\*S\in \mathbb{Z}^{n\times N}_q$ uniformly at random and create $E \in \mathbb{Z}^{(m+1)\times N}_q$ by sampling each entry of it from $D_{\mathbb{Z}_q,\beta_{init}}$. Output $\*A'\*S+\*E+\mu \*G\in \mathbb{Z}^{(m+1)\times N}_q$.

\item[$\bullet$] \tb{MHE.Eval}($C_0,C_1$): To apply the NAND gate, on input $C_0, C_1$, output $G-C_0\cdot G^{-1}(C_1)$.

\item[$\bullet$] \tb{MHE.Dec}$_{sk}$($C$): Let $c$ be column $N$ of $C\in\Z_{q}^{(m+1)\times N}$, compute $b'=sk^{T}c\in\mathbb{Z}_{q}$. Output 0 if $b'$ is closer to 0 than to $\frac{q}{2} \bmod q$, otherwise output 1.

\item[$\bullet$] \tb{MHE.Convert}($C$): Extract column $N$ of $C$.
\end{itemize}

\tbl{Encrypted-CNOT operation.} Given only an MHE-encrypted 1-bit $s\in\{0,1\}$, encrypted-CNOT allows one to perform the encrypted-CNOT operations CNOT$^s$, eventually leading to a QFHE scheme with
classical key generation and classical communication, cf. \cite{mahadev2018classical}. Here, we introduce a generalized version called encrypted-CROT, which allows multi-bits in the encrypted form as control, and supports encrypted control of arbitrary unitary operators, beyond the CNOT gate.
\begin{lem}[Encrypted conditional rotation \cite{ma2022quantum}]\label{12}
Let angle $\alpha\in[0,1)$ be represented in $n$-bit binary form as $\alpha=\displaystyle\sum^{n}_{j=1}2^{-j}\alpha_{j}$ for $\alpha_{j}\in\{0,1\}$. Let $pk_i$ be the public key with trapdoor $t_i$ generated by \textnormal{MHE.Keygen} for $1\leq i\leq n$. Suppose the encrypted trapdoor $\textnormal{MHE.Enc}_{pk_{j+1}}(t_{j})$ is public for $1\leq j\leq n-1$. Given the bitwise encrypted angle $\textnormal{MHE.Enc}_{pk_1}(\alpha)$ and a single-qubit state $|k\rangle$, one can efficiently prepare a ciphertext $\textnormal{MHE.Enc}_{pk_{n}}(d)$, where random parameter $d\in\{0,1\}$, and a state within $\lambda$-negligible trace distance to
\begin{equation}
Z^{d}R_{\alpha}|k\rangle, \tn{  where }   R_{\alpha}=\begin{bmatrix}1 & \\ & e^{2\mi\pi\alpha}\end{bmatrix}
\end{equation}
\end{lem}
\tb{Remark:} In the original version of lemma (cf. Theorem 3.3  in \cite{ma2022quantum}), the output state is $Z^{d}R^{-1}_{\alpha}|k\rangle$. The variant presented here is derived by applying the original lemma to the encrypted shifted angle $(1-\alpha\mod 1)$, and utilizing the relationship $R_\alpha=R^{-1}_{1-\alpha}$.

There is ongoing interest in developing QFHE with classical clients under various security assumptions and models. For example, Brakerski \cite{brakerski2018quantum} improved the security of QFHE schemes by reducing the LWE assumption from a super-polynomial noise ratio (in Mahadev's QFHE) to a polynomial noise ratio, matching the gold standard for LWE-based security; Gupte and Vaikuntanathan \cite{gupte2024construct} further extended QFHE construction to support any underlying FHE scheme, rather than being restricted to the specific GSW matrix-style ones, assuming the existence of dual-mode trapdoor function family; Zhang \cite{zhang2021succinct} avoided trapdoor assumptions in QFHE construction, using only the random oracle model in Minicrypt, at the cost of requiring client to perform quantum computations that are polynomial in the security parameter but still independent of the evaluated circuit depth.

\subsection{Hybrid Quantum-classical Private Information Retrieval}
We modify the definition of single-server private information retrieval from \cite{corrigan2020private} to accommodate a hybrid setting with classical clients and quantum servers. The modification focuses on restricting key generation and communication to be classical.

\begin{defn}[Hybrid quantum-classical private information retrieval]
An $1$-round, $n$-bit Hybrid quantum-classical private information retrieval (QCPIR) is a two-party protocol $\Pi$ between a classical client and a quantum server. The server holds a database $\x\in\{0,1\}^n$. The client holds a index $i\in[n]$. The protocol contains:
\begin{itemize}
   \setlength{\leftskip}{0pt}
\item[$\bullet$] \tn{\textbf{QCPIR.KeyGen}}$((1^{\lambda},n) \rightarrow (ek,dk)):$ the algorithm that takes in security parameter $\lambda$ and database size $n$ and outputs a encryption key $ek$ and a decryption key $dk$.
\setlength{\itemsep}{7pt}

\item[$\bullet$] \tn{\textbf{QCPIR.Query}}$((ek,i)\rightarrow q):$ the classical algorithm that takes in the client's key $ek$ and an index $i\in[n]$, and outputs a query $q$.

\item \tn{\textbf{QCPIR.Answer}}$((q)\rightarrow a):$ the quantum algorithm that takes as input a query $q$, and outputs a classical string $a$.

\item \tn{\textbf{QCPIR.Reconstrct}}$((dk,a)\rightarrow x_i):$ the classical algorithm that takes as input the answer $a$ and key $dk$, and outputs a bit $x_i$.

\end{itemize}
\end{defn}

\textbf{Correctness}. We call $\Pi$ is $\epsilon$-correct if for every $\lambda, n \in\Z$, $x\in\{0,1\}^n$, $i\in[n]$, it holds that
$$\tn{Pr[Reconstrct}(dk,a)=x_i]\geq1-\epsilon,$$
where the probability is taken over any randomness used by the algorithms.

\textbf{Security}. For $\lambda,n\in\Z$, and $i,j\in[n]$, define the distribution
 \begin{equation}
D_{\lambda,n,i}=\left\{               
  \begin{array}{cc}
\quad & ek \leftarrow \tn{KeyGen}(1^{\lambda},n) \\
 \quad & q \leftarrow \tn{Query}(ek,i) \\
  \end{array}
\right\}.
\end{equation}
We call $\Pi$ is correct if for any efficient quantum adversary $\mathcal{A}$, the adversary's advantage satisfies
\begin{align}
\tn{PIRadv}[\mathcal{A},\Pi](\lambda,n):=\max_{i,j}\{\tn{Pr}[\mathcal{A}(1^\lambda,D_{\lambda,n,i})=1]-[\mathcal{A}(1^\lambda,D_{\lambda,n,j})=1]\}<\epsilon
\end{align}

\subsection{Quantum Random Access Memory}

\textbf{Quantum database state.} For a classical database (vector) $\vx:=(x_1,...,x_N)\in \Z_{2^l}^{N}$, where each entry is of binary form $x_i=\sum_{j\in[l]}2^j x^j_i$, the quantum database state that encodes $\vx$ in the computational basis is defined as follows \begin{align}
\frac{1}{\sqrt{N}}\sum_{i\in[N]}\bk{i}\bk{x^{0}_i,x^{1}_i,...,x^{j}_i}.
\end{align}
\textbf{Database preparation unitary}. The quantum database preparation unitary for database $\vx$ is
\begin{align}U:\bk{i}\bk{0} \longrightarrow \bk{i}\bk{x_i}  \tn{  for }  i\in[N]. \label{60}
\end{align}
\textbf{QRAM}. Quantum random access memory (QRAM) is an algorithm that, inspired by the parallel structure of classical RAM, allows to accelerate the preparation of quantum vector states using ancilla qubits (referred to as quantum memory). Details are as follows.


\begin{lem}\emph{(Hybrid quantum random access memory circuit \cite{ giovannetti2008quantum,hann2021resilience})}\label{41}
Given a classical description of an arbitrary $N$-bit database $\vx\in\{0,1\}^{N}$, $M$ ancilla qubits, single and two qubit gates, the preparation unitary operator (\ref{60}) for database $\vx$ can be deterministically implemented with a circuit depth $\bigO{ \frac{N}{M}\log N }$ and $\bigO{M + \log N}$ qubits.
\end{lem}
QRAM captures a tradeoff between time and space complexity. Setting $M=N$ in Lemma \ref{41} results in a state preparation time logarithmic in $N$, using $O(N)$ qubits. Conversely, setting $M$ as a constant increases the preparation time to $O(N)$, while reducing the number of qubits to log$\ N$. The latter case is similar to the cost of a simple state preparation using $N$ controlled gates, where each gate inputs an entry of the vector $\vx$.

A gate-level quantum circuit for implementing QRAM is provided in  \cite[Fig. 9]{hann2021resilience}.

\section{Private Information Retrieve with Classical Client and Single Quantum Server.}

\hspace{0.6cm}\textbf{Binary vector for integer.} For an $n$-bit integer $s\in [2^n]$, we denote its corresponding $n$-dimensional binary vector by the bold letter $\vs\in\{0,1\}^{n}$.

\tb{XOR operation.} The XOR operation (or modulo 2 addition) on two bits $a,b\in\{0,1\}$ is defined by $a \bigoplus b=a+b \mod 2$. For two vectors $\va,\vb\in\{0,1\}^n$, the notation $\va\bigoplus \vb$ denotes bitwise XOR operation. For an integer $a\in[2^n]$ and a vector $\vb\in\{0,1\}^n$, $a \bigoplus \vb$ represents the bitwise XOR between the binary bits of $a$ and binary entries of $\vb$.

\textbf{Parallel quantum gates.} For an $n$-dimensional vector $\vs=(s_0,...,s_{n-1})\in\{0,1\}^n$, and an $n$-qubit state $\bk{\psi}$, the notation $X^{\vs}\bk{\psi}$ means applying the $X^{s_i}$ gate to the $i$-th qubit of $\bk{\psi}$ for each $i\in[n]$.

Following the idea of achieving PIR through homomorphic computations on encrypted indices, we propose a hybrid quantum-classical PIR method. This protocol can run at fastest in time polylog$(N)$ for searching an arbitrary database of size $N$, without relying on any oracle assumption. The basic framework of the protocol is given below \footnote{For convenience, we adopt a concrete scheme from \cite{mahadev2018classical} to construct QCPIR, although any QFHE scheme could serve this purpose. }, and the core operations executed by the quantum server are detailed in Algorithm \ref{20}.

\begin{grayenumerate}

\textbf{QCPIR protocol based on QFHE scheme of \cite{mahadev2018classical}}

\item[$\bullet$]  \textbf{QCPIR.KeyGen.} Run the key generation procedure of MHE scheme.

\item[$\bullet$] \textbf{QCPIR.Query.} The client computes the bitwise encrypted index MHE.Enc$(s)$, where $s=\sum_{i\in[n]}2^is_i$ and $s_i\in\{0,1\}$, and then sends encryptions MHE.Enc$(s)$ to the server.

\item[$\bullet$] \textbf{QCPIR.Anwser.} The server applies the Algorithm \ref{20} to obtain an OTP encryption $m':=\tn{DB}_s \bigoplus a'_2$, together with the corresponding encrypted OTP-key MHE.Enc($a'_2$), and then sends these encryptions to the client.

\item[$\bullet$] \textbf{QCPIR.Reconstrct.} The client decrypts $\tn{MHE.Enc}(a'_2)$ to obtain the OTP-key $a'_2$, followed by computing DB$_s=m'\bigoplus a'_2$.
\end{grayenumerate}

\vspace{0.5cm}
    \begin{breakablealgorithm}
        \caption{Private Information Retrieval with Quantum Server}\label{20}
        \begin{algorithmic}[1] %
            \Statex \hspace{-\algorithmicindent} \textbf{Server-side Input:}  A bitwise encrypted index MHE.Enc$(s)$ where $s=\sum_{i\in[n]}2^is_i$ and $s_i\in\{0,1\}$; the classical description of database DB$_j\in\{0,1\}$ for $j\in[N]$.
            \Statex \hspace{-\algorithmicindent} \textbf{Server-side Output:} The OTP encryption of data DB$_s$, $m':=\tn{DB}_s \bigoplus a'_2$,
           the encrypted OTP-key MHE.Enc($a'_2$).
            \Statex \hspace{-\algorithmicindent} \hrulefill
           \State The server creates an initial state $\bk{0}$, which can be rewritten as $(X^{s_0}\bigotimes\cdots X^{s_{n-1}}\bk{s})\bk{0}$, and shortly denoted by $(X^{\vec{s}}\bk{s})\bk{0}$.
           \State Let the unitary operator $U:=\bk{j}\bk{0}\rightarrow\bk{j}\bk{\tn{DB}_j}$. The server homomorphically compute $U$ to produce a state\footnote{The notation QFHE.Eval$_{Mah}$ specifically denotes the evaluation algorithm of Mahadev's Pauli-OTP-based QFHE scheme, which takes the classical discription of unitary $U$, encrypted Pauli key \(\mathsf{MHE.Enc}(s)\) together with the Pauli-encrypted quantum state as inputs.}
          \begin{align}\tn{QFHE.Eval$_{Mah}$}&(U,\tn{MHE.Enc}(s),X^{(\vec{s},0)}Z^{\vec{0}}\bk{s,0})=X^{\va'}Z^{\vb'}(U\bk{s}\bk{0})\nonumber\\
          &=X^{(\va'_1||a'_2)}Z^{\vb'}(\bk{s}\bk{\tn{DB}_s}_G),\label{102}\end{align}
           together with the ciphertext MHE.Enc$(\va',\vb')$ where $\va'=(\va'_1||a'_2)\in \{0,1\}^n\times \{0,1\}$.
          \State The server measures the register $G$ in (\ref{102}) to obtain a result $m':=\tn{DB}_s\bigoplus a'_2$. 
                   \end{algorithmic}
    \end{breakablealgorithm}

Algorithm \ref{20} has two selectable components: the choice of unitary $U$ for database preparation in Step 3, and the choice of QFHE schemes. These options increase the flexibility of the constructed QPIR protocol, making it better suited to diverse scenarios.

When dealing with general database, choosing the scheme of \cite{mahadev2018classical} as the QFHE scheme, together with QRAM circuit for database preparation as $U$, can achieve a time complexity logarithmic in $N$, as formally stated in the following Proposition \ref{11}. Additionally, choosing other schemes as the underlying QFHE scheme can serve different purposes, such as for a stronger security (by the scheme of \cite{brakerski2018quantum}), or a lower space complexity (by the scheme of \cite{broadbent2009universal}, at the cost of additional client-side quantum computations).

When dealing with database of certain structures, there may exist more efficient quantum implementations, than the general QRAM-based implementations. For instance, if the database entries DB$_i$ is an efficiently computable classical function of $i\in[N]$, then the preparation unitary $\bk{i}\bk{0}\rightarrow\bk{i}\bk{DB_i}$ can be realized more efficiently by simulating classical computing in the quantum computational basis (like operation (5) in \cite{grover2002creating}. This approach, unlike the QRAM method that requiring many ancilla qubits, maintains polylog$(N)$ space complexity.

\begin{prop}\label{11} Given a classical database $DB_i \in \{0,1\}^N$ for $i\in [N]$, a Pauli OTP-encrypted state $X^a Z^b(\sum_{i\in[N]} p_i\bk{i})$ and the encrypted Pauli-keys MHE.Enc$(a,b)$, there exists a quantum algorithm with a circuit depth $\pol{N}$ and $O(N)$ ancilla qubits, that outputs the OTP-encrypted data $X^{\va'} Z^{\vb'} (\sum_{i\in[N]} p_i\bk{i}\bk{DB_i})$ and the encrypted Pauli-keys MHE.Enc$(\va',\vb')$, where random parameters $\va',\vb'\in\{0,1\}^{log N+1}$.
\end{prop}

This proposition follows immediately from QCPIR protocol Algorithm \ref{20}, where the unitary $U$ is implemented using the QRAM circuits in Lemma \ref{41}.







\section{Quantum Fast Implementation of Bootstrapping}

To clarify parameters, we use the following notation:

\begin{itemize}
\setlength{\itemsep}{-3pt}
 \item  \textbf{LWE ciphertext}: LWE$_{Q,L,n}(m)=(\va,\va\cdot\vs+m\frac{Q}{L}+e)\in \mathbb{Z}^{n+1}_{Q}$, where
\item[-]  $Q \in \mathbb{Z}$: modulus of the LWE ciphertext.
\item[-] $L \in \mathbb{Z}$: plaintext space, with $L = 2^l$.
\item[-]$n \in \mathbb{Z}$: dimension of the ciphertext.
\vspace{-0.1cm}
\hspace{-15pt} \item  \textbf{MHE ciphertext}: MHE$_{Q',n'}(\mu)=A'S+E+\mu G\in \Z_{Q'}^{(n'+1)\times (n'+1)\log Q'}$, where
\item[-]$Q'\in\Z$: modulus of MHE ciphertext.
\item[-]$n'\in\Z$: dimension of ciphertext.
\vspace{-0.1cm}
\item  \textbf{Bootstrapping parameters}:
\item[-]$N_\star\in\Z$: amplitude scaling parameter, which will impact the success probability of the Algorithm \ref{123}. Denote $N_{\star}=2^{n_{\star}} $.
\item[-]$L'\in\Z$: bit-size of output message. Let $L=2^{l'}$.
\end{itemize}

\vspace{-0.2cm}

For simplicity, we assume that all the above parameters (e.g., ciphertext modulus $Q$, dimension $n$, and plaintext space $L$) are powers of $2$. This assumption, commonly made in the literature on FHE, cf. \cite{chillotti2020tfhe}, facilitates digital realization.

Subscript parameters in ciphertexts are sometimes omitted when clear from the context.

\textbf{Imaginary unit.} We use $I$ to denote the imaginary unit throughout this section.\\

In this section, we first present a new technique called quantum blind rotation, and then apply it to construct quantum fast bootstrapping and functional bootstrapping. An overview of the process for fast bootstrapping and functional bootstrapping is given in Figure \ref{6}.

\subsection{Quantum Blind Rotation}\label{4.1}

At a high level, \textbf{quantum blind rotation} involves simulating the decryption process within the amplitudes of qubits, followed by translating the decrypted message in amplitudes to the computational basis via the quantum Fourier transform, also known as phase estimation. However, to ensure privacy, all these procedures must be executed in the encrypted environment.

Specifically, given an LWE ciphertext $(\va,b:=\va\cdot \vs+m+e)$ and the encrypted key MHE.Enc$(s)$, our first goal is to homomorphically compute the LWE-phase $(b-\va\cdot\vs)$ at qubits amplitudes. Although the value of $s$ is unknown, Lemma \ref{12} allows us to perform the encrypted private-key MHE.Enc$(\vs)$ controlled rotation $R_{\va}$. This results in a quantum state, where the LWE-phase $(b-\va\cdot\vs)$ is encoded in the phase factor $e^{2\pi I (b-\va\cdot\vs)y}$ of the $y$-th computation basis state, up to a Pauli-mask whose keys are given in encrypted form.

Fortunately, the impact of Pauli masks is ignorable during quantum homomorphic computations. Therefore, we can continue to perform homomorphic QFT on the resulting state to translate the LWE-ciphertext phase onto computational basis within a certain precision. Ultimately, this process yields an approximation to the message $m$ in OTP-encrypted form, together with OTP-keys in MHE-encrypted form. Details are as follows.

    \begin{breakablealgorithm}
        \caption{Quantum Implementation of Blind Rotation}\label{123}
        \begin{algorithmic}[1] 
            \Require An LWE ciphertext $(\va,b):=(\va,\va\cdot\vs+m\frac{Q}{L}+e)\in \mathbb{Z}^{n+1}_Q$ where $\vs= (s_i)_{i\in[n]} \in\{0,1\}^{n}$, encrypted secret keys MHE.Enc$(s_i)$, encrypted trapdoor $\textnormal{MHE.Enc}_{pk_{j+1}}(t_{j})$ for $1\leq j\leq n_\star-1$; bootstrapping parameters $N_\star=2^{n_\star}$ and $L'=2^{l'}$.
            \Ensure The OTP-encryption $c:=\lfloor L' \frac{m}{L}\rceil  \bigoplus \vec{d_1}$ where $\vec{d_1}\in\{0,1\}^{l'}$, encrypted OTP-key
             $\textnormal{MHE.Enc}_{pk_{n_\star}}(\vec{d_1})$.
           \Statex \hspace{-0.5cm}0:  Let $\va=(a_i)_{i\in[n]}\in\Z^{n}_Q$.
\Statex  Define $a'_i=\rd{\frac{N_\star}{Q}a_i}\in[N_\star]$, and denote $\va'=(a'_i)_{i\in[n]}$.
\Statex  Express $a'_i$ in binary form as $a'_i=\sum_{j\in[n_\star]}2^j a'_{i,j} \tn{ where } a'_{i,j}\in\{0,1\}.$
\Statex  The fraction $\alpha^{(i)} := \frac{a'_i s_i}{N_\star} \in [0,1]$ is represented in the $n_\star$-bit binary form as
$\alpha^{(i)}=\sum_{j\in[n_\star]}2^{j-n_{\star}}\alpha^{(i)}_{j}$ \tn{ where }$\alpha^{(i)}_{j}=a'_{i,j}s_i\in\{0,1\}$.
           \State Prepare bit-wise MHE encryptions of $\alpha^{(i)}$, by computing MHE.Enc$_{pk_1}(\alpha^{(i)}_j):=a'_{i,j}\times$ MHE.Enc$_{pk_1}(s_i)$ for $j\in[n_\star], i\in[n]$. With these encryptions, Lemma \ref{12} allows to perform MHE.Enc$_{pk_1}(\alpha^{(i)})$-controlled rotation $R_{\alpha^{(i)}}$ on the initial state $\bk{+}$ to obtain a state
           \begin{align}
           \frac{1}{\sqrt{2}}Z^{w_i}\sum_{y\in[2]}\epi{\alpha^{(i)} y}\bk{y},\label{14}
            \end{align}
           together with the encrypted Pauli-Z key MHE.Enc$_{pk_{n_\star}}(w_i)$, where $w_i\in\{0,1\}$.
            \State For $i\in[n]$, successively apply MHE.Enc$_{pk_1}(\alpha^{(i)})$-controlled phase rotation on state $\bk{+}$. The result is
            \begin{align}Z^{w'} \sum_{y\in[2]} \epi{\frac{\vec{a'} \cdot \vec{s}}{N_\star} y}\bk{y},\label{13}
            \end{align}
            where the Pauli-Z key $w':=\sum_{i\in[n]}w_i \mod 2$, and its encryption MHE.Enc$_{pk_{n_\star}}(w')$ can be produced by homomorphic additions of MHE.Enc$_{pk_{n_\star}}(w_i)$.
             \State Set $b'=\frac{N_\star}{Q} b$. Apply phase rotation $R_{b'}:=\begin{bmatrix*}[r]1 &\ \\  &\ \epi{\frac{b'}{N_\star}} \end{bmatrix*}$ on  (\ref{13}) to get state
               \begin{align}Z^{w'} \sum_{y\in[2]} \epi{\frac{b'-\vec{a'} \cdot \vec{s}}{N_\star} y}\bk{y}. \label{15} \end{align}
           \State Repeat $l'$ times steps $2-3$ to produce the Pauli-OTP encryption of the following $l'-$qubit state
            \begin{align}
            &\frac{1}{2^{l/2}}(\sum_{y_0\in[2]}e^{2\pi I \frac{b'-\vec{a'} \cdot \vec{s}}{N_\star} y_0}\bk{y_0})(\sum_{y_1\in[2]}e^{2\pi I \frac{b'-\vec{a'} \cdot \vec{s}}{N_\star} 2 y_1}\bk{y_1})
            \cdots (\sum_{y_{l'-1}\in[2]}e^{2\pi I \frac{b'-\vec{a'} \cdot \vec{s}}{N_\star} 2^{l'-1} y_{l'-1}}\bk{y_{{l'}-1}})\nonumber\\
            &=\frac{1}{2^{l/2}}\sum_{y\in[L']}e^{2\pi I \frac{b'- \vec{a'} \cdot \vec{s}}{N_\star}  y}\bk{y},\label{2}
            \end{align}
            as well as the encrypted Pauli-Z keys $\textnormal{MHE.Enc}_{pk_{n_\star}}(\vec{d})$, where $\vec{d}\in\{0,1\}^{l'}$.

           \State Homomorphically evaluate the $l'-$qubit quantum Fourier transform on state (\ref{2}) to get a Pauli-OTP encrypted state
           \begin{align}\label{1}
           X^{\vec{d_1}}Z^{\vec{d_2}}\left( \frac{1}{L'}\sum_{k\in[L']} (\sum_{y\in[L']}e^{2\pi I (\frac{b'-\vec{a'}\cdot\vec{s}}{N_\star}-\frac{k}{L'})y})\bk{k}  \right),
            \end{align}
            as well as the encrypted Pauli-keys $\textnormal{MHE.Enc}_{pk_{n_{\star}}}(\vec{d_1},\vec{d_2})$,\footnote{In fact, during the homomorphic computation of $l'$-qubit QFT within negl($k$) precision, additional poly$(l',k)$ MHE keyswitch is required. We temporarily ignore the precision issue, and simply use $n_{\star}$ in place of $n_{\star}$+poly$(l',k)$. Further discussion is deferred to the efficiency analysis part.} where $\vec{d_1}, \vec{d_2}\in\{0,1\}^{l'}$
          \State Measure the state of (\ref{1}). Return the measurement outcome $c\in\{0,1\}^{l'}$ and encrypted Pauli-X keys $\textnormal{MHE.Enc}_{pk_{n_\star}}(\vec{d_1})$.
                   \end{algorithmic}
    \end{breakablealgorithm}

Next, we show that the Algorithm \ref{123}, with a high probability, outputs the scaled message $\lfloor L' \frac{m}{L} \rceil$ in OTP-encrypted form, under certain parameter regimes.

In Algorithm \ref{123}, Steps 1 to 3 simulate the computation of the LWE-phase $b-(\va\cdot\vs) \mod Q$ using the amplitude of a single-qubit. In more detail, Step 1 prepares encryptions for $a'_is_i$ where $i\in[n]$. Step 2 realizes the summation $\sum_{i\in[n]}a'_is_i$ using $n$ number of encrypted-CROTs, where each encrypted-CROT introduces a phase factor $e^{2\pi I \frac{a'_i s_i}{N_\star} }$, thereby leading to a cumulative phase in the exponent. In Step 3, after performing $R_{b'}$ to add a phase factor $e^{2\pi I \frac{b'}{N_\star} }$, the LWE-phase is computed at a single-qubit's amplitude, up to a Pauli mask, as shown in (\ref{15}).

Step 4 expands the 1-qubit state (\ref{15}) to an $l'$-qubit state as shown in (\ref{2}). This expansion is done qubit by qubit: for each subsequent qubit, an encrypted-CROT with a doubled rotation angle is applied, using the encrypted fractional bits obtained from Step 1. The resulting state (\ref{2}) is of a standard input form for QFT. After homomorphically performing $l'$-qubit QFT on this state, we will obtain a state (\ref{1}), which actually approximates $ \bk{ \lfloor \frac{b-\va\cdot\vs}{Q}L'  \rceil}$, up to Pauli masks.

To clarify the final outcomes after measuring the state (\ref{1}), we present the following Propositions \ref{220} and \ref{21}.\\

\tbl{Success probability of Algorithm \ref{123}.} We conduct the discussion in two cases: $L'\geq L$ and $L'<L$. For first case of $L'\geq L$, Proposition \ref{220} states that Algorithm \ref{123} outputs\footnote{Recall that plaintext sizes $L,L'$ are both powers of $2$.} $\lfloor L' \frac{m}{L}\rceil=m\frac{L'}{L}$ with a high probability, under a reasonable choice of parameters.

\begin{prop}[Quantum blind rotation for extended plaintext] \label{220}
For powers of two $L$ and $L'$ with $L' \geq L$, Algorithm 2 outputs an OTP encryption of $m\frac{L'}{L}$, as well as the encrypted OTP-keys, with a probability of success $p\geq 1-(L'B/Q + L'n/2N_\star)^4\pi^4/3$, where $B$ is the error bound of input LWE ciphertext in Algorithm 2,
\end{prop}

\textbf{Remark:}{ For practical LWE parameters $Q=2^{29}$, $L=2^{14}$, $B=2^6$, $n=2^{10}$, quantum blind rotation (Algorithm \ref{123}) has a high success probability of $p>1-2^{-30}$ by setting $L'=L$ and $N_\star=2^{29}$. Notably, the last term on the RHS of inequality (\ref{5}) results from rounding, and will vanish when $Q$ divides $N_\star$. In the asymptotic case, we can set LWE parameters $Q=\Omega(n^2)$, $B=O(n^{0.5})$, $L=n$, and bootstrapping parameters $N_\star=\Omega(n^{2.5})$, $L'=n$ to guarantee a probability $p> 1- O(1/n^{0.5})$.} 
\begin{proof}
In the step 6 of the algorithm, we first estimate the probability of obtaining the measurement result $k\oplus \vec{d_1}$ for arbitrary $k \in [L']$. Denote $\tr:=\frac{k}{L'}-\frac{b'-\vec{a'}\cdot\vec{s}}{N_\star}$. Then, the probability is given by
\begin{align}\label{3}
p&=|\frac{1}{L'}\sum_{y\in[L']}\epi{\tr y}|^2= |\frac{1}{L'}\frac{1-\epi{\tr L'}}{1-\epi{\tr }}|^2=\frac{1}{L'^2}|\frac{\sin \pi \tr L'  }{\sin \pi \tr }|^2
\end{align}
Note that the function $\frac{ \sin x}{x}$ is decreasing for $x \in [0,\pi/2]$. So, for a constant $\epsilon\in(0,1/2)$, when $|\tr|<\epsilon/L'$ we have
$$ |\sin \tr L' \pi | \geq  |\tr L' \pi  \frac{\sin \epsilon\pi}{\epsilon\pi}| \textnormal{ \qquad and \qquad} |\sin \tr \pi| \leq |\tr \pi|,$$
Consequently, the lower bound of the probability in (\ref{3}) is given by
\begin{align}\label{4}
p\geq |\frac{\sin \epsilon \pi }{ \epsilon \pi}|^2 > 1- \frac{(\epsilon \pi)^4}{3} ,
\end{align}
where the last inequality follows from that $\sin x>x- x^3 / 6$ (for any $ x > 0$).

To further determine $\epsilon$, notice that

\begin{align}
|\frac{k}{L'}-\frac{b'-\vec{a'}\cdot\vec{s}}{N_\star}|&=|\frac{k}{L'}-\frac{b'-\sum_{i\in[n]}\rd{ N_\star (a_i/Q)}s_i}{N_\star}| \label{22}\\
&=\left|\frac{k}{L'}- \frac{b-\sum\limits_{i\in[n]}a_is_i}{Q}+  \frac{\sum\limits_{i\in[n]} N_\star \left(\frac{a_i}{Q}\right) s_i-\rd{ N_\star \left(\frac{a_i}{Q}\right)}s_i}{N_\star}\right| \\
&\leq |\frac{k}{L'}-\frac{m}{L}|+|\frac{e}{Q}|+\frac{n}{2N_\star} \label{5}
\end{align}

Since $L|L'$, then (\ref{5}) implies that
\begin{align}|\tr|<|\frac{k}{L'}-\frac{m}{L}|+(\frac{B}{Q}L'+L'\frac{n}{2N_\star})/L'.\end{align}
By setting $\epsilon=L'B/Q + L'n/2N_\star$ in (\ref{4}), the probability of obtaining the OTP-encrypted value $k=m L'/L$ as the measurement outcome is at least $1-(L'B/Q + L'n/2N_\star)^4\pi^4/3$.
\end{proof}

With all other parameters fixed, setting $L'=L$ can maximize the success probability. It also achieves optimal efficiency, since choosing a larger $L'$ only produces an encryption of $m$ padded with additional zeros, which provides no useful information but increases complexity.

Proposition \ref{21} states that in the case of $L'<L$, Algorithm \ref{123} outputs the integer closest to $L'\frac{m}{L}$ with a probability at least $p>\frac{4}{\pi^2}$. To simplify the discussion, the proposition assumes a reasonable parameter regime, while the conclusion of a constant lower bound holds for more broader parameter settings.

\begin{prop}[Quantum blind rotation for compressed plaintext] \label{21}
When $L'<L$, $N_\star \gg N$, and $B/Q \ll m/L $, the probability of the algorithm outputs an OTP-encryption of $\lfloor m\frac{L'}{L} \rceil$ is at least $\frac{4}{\pi^2} (>0.405) $.
\end{prop}

\begin{proof}
We use the notations from the immediately above proof. Given that $\frac{B}{Q} \ll\frac{m}{L}$, we treat the LWE-phase $\frac{b-\va\cdot\vs \mod Q}{Q}$ as $\frac{m}{L}\in[0,1]$ throughout this proof. Consider the $L'$-division points in the interval $[0,1]$. For $\frac{m}{L}$ that is not an $L'$-division point, there must be two closest adjacent $L'$-division points, denoted by $\frac{k_1}{L'}$ and $\frac{k_2}{L'}$ where $k_1,k_2\in[L']$. Without loss of generality, assume $k_1 =\lfloor L' \frac{m}{L} \rfloor$, and define the distance $h_1= |\frac{k_1}{L'}-\frac{ m}{L}|$, so that the distance $h_2=|\frac{k_2}{L'}-\frac{m}{L}|$ satisfies $h_2+h_1=\frac{1}{L'}$.

By arguments similar to $(\ref{3})\sim(\ref{4})$, after measuring the state (\ref{1}) in step 6, the probabilities of observing $k_1$ and $k_2$ is

\begin{align}
\frac{1}{L'^2}\left( \left|\frac{\sin (\pi h_1 L')  }{\sin (\pi h_1) }\right|^2+\left|\frac{\sin (\pi h_2 L')  }{\sin (\pi h_2) }\right|^2 \right)\geq &\frac{2}{L'^2}\left|\frac{\sin (\pi h_1 )L' \sin (\pi h_2 L')  }{\sin (\pi h_1) \sin( \pi h_2)} \right|\\
=&\frac{2}{L'^2} \frac{\cos(\pi (h_1+h_2)L') - \cos(\pi (h_1-h_2)L')}{\cos\pi (h_1+h_2) - \cos\pi (h_1-h_2) }\label{24} \\
\geq& \frac{2}{L'^2}  \left| \frac{1}{\sin\left(\frac{\pi}{2L'}\right)} \right|^2   \qquad \forall h_1\in (0, \frac{1}{2L'})  \\
\geq& \frac{8}{\pi^2}   \qquad \qquad  \qquad  \qquad \forall L' \in \Z_{+}
\label{23}
\end{align}

The function $f(h_1):=\frac{\sin \pi h_1 L'  }{\sin \pi h_1 }$ is decreasing on the interval $h_1\in [0,\frac{1}{2L'}]$. So, the probability of measuring a outcome $\lfloor  m\frac{L'}{L} \rceil$, which is closer to
$ m\frac{L'}{L}$ and serves as input that yields a higher function value, is at least $ \frac{4}{\pi^2}$.

For the case where $\frac{m}{L}$ is just an $L'$-division point, the proof proceeds similarly.
\end{proof}

\tbl{Quantum amplitude-basis joint encoding.} By substituting $h_2=1/L'-h_1$ into the RHS of the equality in (\ref{24}), it becomes monotonically decreasing with respect to $h_1$. This implies that when $L'<L$, in addition to the first $l'$ most significant bits of plaintext $m$ being directly represented in the computational basis state of (\ref{1}), the lower bits of $m$ are also represented in the amplitudes of (\ref{1}). Thus, by multiple measurements of the state of (\ref{1}), it is possible to recover $m$ with a precision greater than $l'$ bits.

Interestingly, this suggests that $l'$ qubits, combined with their qubits' amplitudes (determined by multiple measurements), can represent a message of size $l>l'$ (as in Algorithm \ref{123}). While this encoding method may not help reduce the overall time complexity, it could shorten the time for individual algorithm executions, at the cost of increasing the number of executions.

\subsection{Quantum Implementation of Bootstrapping}\label{101}

To realize quantum bootstrapping, we first execute Algorithm \ref{123} with setting $L'=L$. Then, the resulting outputs can be combined into an encryption LWE.Enc$_{Q',2^{l},n'}(m)$. Finally, key switching is employed to adjust both the dimension and modulus into the desired form, LWE.Enc$_{Q,2^{l},n}(m)$. The detailed subroutines involved are as follows.\\

\tbl{Combining the outputs to an LWE ciphertext.} Recall that the outputs of Algorithm \ref{123} include an OTP encryption $c := m'\bigoplus \vec{d_1} \in \{0,1\}^{l'}$, together with MHE-encrypted OTP-keys MHE.Enc$(\vec{d}_1)$. Here, $m'$ is the $l'$-bit scaled plaintext, and is interpreted as a bit string $(m_j)_{j\in[l']}\in\{0,1\}^{l'}$, where $m':=\sum_{j\in[l']} 2^{l'-j-1} m'_j$ and $m'_j\in\{0,1\}$. Let the OTP encryption $c$
consist of bits $c_j:=m'_j \oplus d_{1,j} \in\{0,1\}$ for $j\in[l']$. It remains to show how to convert this OTP encryption into an LWE encryption, with the help of MHE-encrypted OTP keys.

Firstly, the LWE encryption of $d_{1,j}$ is known. To see this, the $k$-th column from right in GSW-style matrix ciphertext MHE.Enc$(d_{1,j})$, denoted as MHE.Enc$(d_{1,j};k)$, is just an LWE encryption LWE.Enc$_{Q',2^k,n'}(d_{1,j}):=(\va,\va\cdot\vs+e+ d_{1,j} \frac{Q'}{2^{k}} )$, for any $j \in[l']$ and $k \in[l']+1$ (where $k\leq l' \leq \log_2 Q'$).

Now with ciphertexts MHE.Enc$(d_{1,j})$ and $c_j$ at hand, and according to the equality
\begin{align}m'_j Q'/2^{k}&=(c_j \oplus d_{1,j}) Q'/2^{k}= (c_j  + d_{1,j})Q'/2^{k} -  c_j d_{1,j} Q'/2^{k-1}, \label{222}\end{align}
one can generate the encrypted 1-bit of $m'$, LWE.Enc$_{Q',Q',n'}(m'_j Q'/2^{k})$, by homomorphically evaluating XOR on MHE.Enc$(d_{1,j};k)$ and LWE.Enc$_{Q',Q',n'}(c_j Q'/2^{k})$. That is done by computing
\
\begin{align}
(\vec{0},c_j Q'/2^{k})&+\textnormal{MHE.Enc}(d_{1,j};k)-c_j\times \textnormal{MHE.Enc}(d_{1,j};k-1).\hfill
\end{align}

Consequently, the LWE encryption of $m'$ follows by homomorphic additions

\begin{align}&\tn{LWE.Enc}_{Q',2^{l'},n'}(m' )=\tn{LWE.Enc}_{Q',Q',n'}(\sum_{j\in[l']} m'_j  Q'/2^{j+1})=\sum_{j\in[l']} \tn{LWE.Enc}_{Q',Q',n'}( m'_j  Q'/2^{j+1}) \label{333}
\end{align}

Given the setting $l'=l$, we now obtain the ciphertext LWE.Enc$_{Q',2^{l},n'}(m)$. Next, we transform it to LWE.Enc$_{Q,2^{l},n}(m)$ using \emph{key switching}, as described below.\\

\tbl{Key switching and noise analysis.} Key switching transforms the private keys, together with the modulus and dimension, into a new form, provided that the encryption of old private keys are available.

We begin with the ciphertext in (\ref{333}), which has the form $\tn{LWE.Enc}_{Q',L,n'}(m):=\left((a_i)_{i\in [n']}, b\right)$, where $a_i$ is represented in binary form as $a_i=\sum_{j\in[\log Q']} 2^{j} a_{i,j}$ for $i\in[n']$. The error bound in the ciphertext is denoted by $Err(\tn{LWE(m)})$, which will be analyzed later. Given the fresh LWE-encrypted keys $\tn{LWE}_{Q,Q'2^{-j},n}(s_j)$, with the fresh noise bound denoted by $Err(\tn{LWE(sk)})$, the key switching operation is to compute
\begin{align}(0, \lfloor b\frac{Q}{Q'} \rceil)\ - \sum_{{i\in[n'],j\in[\log Q']}}a_{i,j }\tn{LWE}_{Q,Q'2^{-j},n}(s_j) \mod Q,\label{51}\end{align}
which approximately homomorphically evaluates $(b-\va \cdot \vs)Q/Q'$ in the new LWE encryption setting. This results in a ciphertext LWE$_{Q,L,n}(m)$, with the noise bounded by
\begin{align}B_f:=Err(\tn{LWE(sk)})n'\log Q'&+Err(\tn{LWE(m)})Q/Q'+\sqrt{n'\log Q'},\label{50}\end{align}
where the last term $\sqrt{n'\log Q'}$ is the heuristic bound for rounding.\footnote{Obviously, the new noise bound $B_f $ is greater than the previous bound $Err(\tn{LWE}(m))$. So, the key switching alone can not be used for the purpose of noise reduction.}

In the primary case discussed in this paper, where $Q'\gg Q$, the dominant error contribution comes from the first term on the RHS of (\ref{50}). This term increases linearly with the number of homomorphic additions, and is the fresh noise bound multiplied by a factor $n' \log Q'$. Furthermore, when $Q'\gg Q$, the terms $\tn{LWE}_{Q,Q'2^{-j},n}(s_j)$ with subscripts $j\leq \log \frac{Q'}{Q}$ are approximately encryptions of $0$, and can be ignored in the summation in (\ref{51}). This leads to a tighter upper bound on the error:
\begin{align}B_f:=Err(\tn{LWE(sk)})n'\log Q&+Err(\tn{LWE(m)})Q/Q'+\sqrt{n'\log Q},\end{align}


Next, we give an upper bound for the error term $ Err(\tn{LWE(m)})$ in equation (\ref{50}) using parameters from MHE schemes.\\

\tbl{Noise bound in MHE.} Let $\beta_{acc}$ be the accumulated noise bound for MHE ciphertext $AS+E+\mu G \in \mathbb{Z}_{Q'}^{(n'+1)\times (n'+1)\log Q'}$, such that $||E||_{\infty}<\beta_{acc}$ throughout homomorphic computations. For the MHE ciphertext output by algorithm \ref{123}, the noise in each of its columns, when viewed as an LWE encryption, is bounded by $(n'+1)\beta_{acc}$. Consequently, the noise in the combined LWE of (\ref{333}), also referred in (\ref{50}), is bounded by
\begin{align} Err(\tn{LWE(m)}):= 2l'(n'+1)\beta_{acc}. \label{53}\end{align} To correctly decrypt (\ref{333}) and recover $m'$, the noise bound $\beta_{acc}$ must satisfy a stricter condition $\beta_{acc}<\frac{Q'}{8L'l'(n'+1)},$ compared to the original MHE scheme's requirement $\beta_{acc}<\frac{Q}{4(n'+1)}$ for achieving QFHE.\\

\tbl{Noise reduction .} To understand how quantum bootstrapping (cf. Figure \ref{6}) may reduce noise, we combine the noise bound after key switching (\ref{50}) with the bound (\ref{53}). This implies that the noise ratio for refreshed ciphertext after bootstrapping is lower bounded by the inverse of
\begin{align}\frac{B_f}{Q}=\frac{\sqrt{n'\log Q'}+Err(\tn{LWE(sk)})n'\log Q'}{Q}
+\frac{2l'(n'+1)\beta_{acc}}{Q'},\hfill\label{81}\end{align}
According to the MHE parameter relations $Q'=\tn{poly}(\lambda)^{\Theta (\log \lambda)}$ and $\frac{\beta_{acc}}{Q'}=\tn{negl}(\lambda)$ (cf. (62) of [Mah18]), the first term on the RHS of (\ref{81}) is dominant. Therefore, the bootstrapping procedure shown in Figure \ref{6} can effectively reduce accumulated noise, when the noise ratio of input LWE$_{Q,L,N}$ falls below $Q/B_f$, especially when the accumulated input noise exceeds $n'\log Q'$ times the fresh noise $Err(\tn{LWE(sk)})$.


\subsection{Quantum Implementation of Functional Bootstrapping}\label{4.3}



%
%


We present the main idea behind the efficient quantum algorithm for functional bootstrapping. Details are provided in Algorithm \ref{909}.

We begin with the outputs of Algorithm \ref{123}. Proposition \ref{220} states that when setting $L'\geq L$ and selecting suitable parameters, the outputs in Step 6 are a Pauli-OTP encryption of the scaled plaintext $ L' \frac{m}{L} $, together with the encrypted Pauli-keys MHE.Enc$(\vec{d_1, \vec{0}})$.

Now, to evaluate a function $f(m) : \Z_{L} \rightarrow \mathbb{Z}_{\tilde{L}}$ over plaintext $m$, we extend the domain and define a new function, called the ``test function" $\tf : \Z_{L'} \rightarrow \Z_{\tL}$, such that $\tf(L'\frac{m}{L})=f(m)$ for all $m\in[L]$. The function $\tf$ is well-defined because $m_1\frac{L'}{L}\neq m_2\frac{L'}{L}$ for any $m_1 \neq m_2 $ when $L'\geq L$. However, the function may not be uniquely defined over the extended domain $[L']$.

Using QFHE schemes, similar to Step 3 of Algorithm \ref{20}, we can homomorphically evaluate $\tf$ on the Pauli-encrypted state $X^{\vec{(d_1||\vec{0})}}\bk{ L' \frac{m}{L}}\bk{\vec{0}}$ and the encrypted Pauli-keys MHE.Enc$(\vec{d_1}, \vec{0})$. This evaluation produces the state $X^{\vec{d'_1}}Z^{\vec{d'_2}}\bk{ L' \frac{m}{L}}\bk{\tf( L' \frac{m}{L})}$, and new keys MHE.Enc$(\vec{d'_1}, \vec{d'_2})$.

After measuring the second register, the classical OTP-encryptions of $f(m)$ are obtained, which can then be converted into the LWE form by the conversion method presented in (\ref{222})$\sim$(\ref{333}).

\vspace{0.5cm}
    \begin{breakablealgorithm}
        \caption{Quantum Implementation of Functional Bootstrapping}\label{909}
        \begin{algorithmic}[1] 
            \Require OTP$_{\vec{d}_1}$ $(m'):=m'\bigoplus \vec{d}_1\in \{ 0,1 \}^{l'}$; encrypted private key MHE.Enc$_{Q',n'}(\vec{d}_1)$; classical function $f(m) : \Z_{L} \mapsto \Z_{\tL}$; $L'\geq L$

            \Ensure LWE$_{Q',\tL,n'}$$(f(m))$
           \State Set $\tf(m') : \Z_{L'} \rightarrow \Z_{\tL}$ such that $\tf(L'\frac{m}{L})=f(m)$ for all $m\in[L]$.
           \State Create $(l'+\tilde{l})$-qubit initial state $\bk{m',\vec{0}}$.
           \State Use QFHE scheme to homomorphically compute $U:\bk{m'}\bk{0}\rightarrow \bk{m'}\bk{f(m')}$ to obtain the state
           \begin{align}
                      \tn{QFHE}(\tilde{f},X^{(\vec{d}_1||\vec{0})},\bk{m',\vec{0}})=X^{(\vec{d'_1}||\vec{d'_2})}Z^{\tilde{\vb}} \bk{m',\tilde{f}(m')}\label{8}
                      \end{align}
           as well as the encrypted key MHE.Enc$_{Q',n'}(\vec{d'_2})$.
          \State Measure the second register of the state (\ref{8}). The result is $c:=f(m)\bigoplus \vec{d'_2}\in\{0,1\}^{\tilde{l}}$.
          \State Use combine-to-LWE methods $(\ref{222})\sim(\ref{333})$ on ciphertexts $c$ and MHE.Enc$_{Q',n'}(\vec{d'_2})$ to prepare LWE$_{Q',\tilde{L},n'}(f(m))$.
                   \end{algorithmic}
    \end{breakablealgorithm}

\textbf{Complexity.} The whole process for quantum functional bootstrapping (Algorithm \ref{123} and Algorithm \ref{909}) consists of four main steps, cf Figure \ref{6}. The complexity is dominated by the number of 1-bit CROT operations, which is calculated as follows:

1. Steps 1-3 (of Algorithm 2) computes the LWE-phase state (\ref{15}) on a single-qubit , requiring $O(n, \log N_{\star})$ 1-bit CROT operations.

2. Preparing the $l'$-qubit state (\ref{2}) for LWE-phase uses O$(\log N_{*}, n, l')$ 1-bit CROTs.

3. Homomorphically performing \(l'\)-QFT within \(\text{negl}(\lambda)\) precision requires O$(\lambda^2)$ homomorphic evaluations of non-Clifford gates (as guaranteed by the optimal Solovay-Kitaev algorithm \cite{dawson2005solovay}), and thus O$(\lambda^2)$ 1-bit CROTs. \footnote{More efficient approaches exist, such as using low $T$-depth QFT circuit \cite{nam2020approximate} or other QFHE method \cite{ma2022quantum}, which can reduce the number of non-Clifford evaluations to O$(\lambda)$. We set parameters $l',n,\log N_{\star}$ are all poly$(\lambda)$, to ensure a negl$(l',n,\log N_{\star})$-precision. }

4. For each output bit of the function $f(m')$, homomorphically evaluating QRAM circuit over \(l'\)-qubit state $\bk{m'}$ of (\ref{8}) involves a circuit of \(O(l')\)-depth controlled-SWAP gates (cf. \cite[Fig. 9]{hann2021resilience}). Each controlled-SWAP gate can be implemented using a Toffoli gate and two CNOT-gates \cite{ns2020}, as shown below:
\begin{align}\text{Controlled-SWAP} = (I \otimes \text{CNOT}_{2,1}) \, \text{Toffoli} \, (I \otimes \text{CNOT}_{2,1}).\end{align}
So, this step requires \(O(l',\tl)\) 1-bit CROTs.

In total, poly$(l',n,\log N_\star,\tl)$ number of 1-bit CROT are used in quantum functional bootstrapping. The cost of 1-bit CROT is comparable to that of encrypted-CNOT in \cite{mahadev2018classical}, and scales polynomially with the dimension $n'$ and the ciphertext-size log $Q'$ for the encrypted control bit \cite{ma2022quantum}. Other operations, like homomorphic evaluations of Clifford gates, are relatively inexpensive and scale polynomial in $l'$.

Therefore, the runtime of functional bootstrapping is poly$(l',n,\log N_\star,\tl)$, providing an exponential improvement over the dependence on the plaintext size, which is $2^{l'}$ for the best-known classical algorithms.

\begin{thm}\label{4.4}
Given an LWE encryption of plaintext $m\in[L]$, an arbitrary function $ f: \Z_{L} \rightarrow \mathbb{Z}_{\tilde{L}}$, there is a quantum algorithm outputs the LWE encryption of $f(m)$ with a runtime of \tn{polylog}$(L,\tilde{L})$ and a qubit cost of O$(L,\log \tilde{L})$ . Moreover, if the function $f$ is efficiently computable in time \tn{polylog}$(L,\tilde{L})$ with space \tn{polylog}$(L,\tilde{L})$, then the qubit cost can also be reduced to \tn{polylog}$(L,\tilde{L})$.
\end{thm}
\begin{proof}
By combining Algorithm \ref{123} and Algorithm \ref{909} with setting $L=L'$, the first part of the theorem follows from using QRAM circuit\footnote{Although QRAM provides polylogarithmic circuit depth using \(O(N)\) quantum memory, its total gate count remains \(O(N\,\mathrm{polylog}\,N)\). Therefore, the claimed improvement concerns parallel time/circuit depth under the QRAM model, rather than total gate complexity.} to implement $U$ in Step 3 of Algorithm \ref{909}. For the second part where $f$ is efficiently computable, the unitary operator $\bk{m'}\bk{0}\rightarrow\bk{m'}\bk{f(m')}$ can be implemented directly by simulating classical computations in quantum computational basis state, with the complexity comparable to that of classical setting.
\end{proof}

 \bibliographystyle{alpha}

\bibliography{apssamp}

\end{document}